\documentclass[twocolumn, times]{aastex631}
\usepackage{amsmath}
\usepackage{soul}
\usepackage{xcolor}
\usepackage[all]{hypcap}

\newcommand{\hx}[1]{\hspace{#1px}}
\def\sfxc{{\sc sfxc}}
\def\casa{{\sc casa}}
\def\aips{{\sc aips}}

\begin{document}

\title{The Radio Parallax of the Crab Pulsar: A First VLBI Measurement Calibrated with Giant Pulses}
\shorttitle{Radio Parallax of the Crab Pulsar}

\author[0000-0003-4530-4254]{Rebecca Lin}
\affil{Department of Astronomy and Astrophysics, University of Toronto, 50 St. George Street, Toronto, ON M5S 3H4, Canada}

\author[0000-0002-5830-8505]{Marten H. van Kerkwijk}
\affil{Department of Astronomy and Astrophysics, University of Toronto, 50 St. George Street, Toronto, ON M5S 3H4, Canada}

\author[0000-0001-6664-8668]{Franz Kirsten}
\affil{Department of Space, Earth and Environment, Chalmers University of Technology, Onsala Space Observatory, 439 92, Onsala, Sweden}

\author[0000-0003-2155-9578]{Ue-Li Pen}
\affil{Institute of Astronomy and Astrophysics, Academia Sinica, Astronomy-Mathematics Building, No. 1, Sec. 4, Roosevelt Road, Taipei 10617, Taiwan}
\affil{Canadian Institute for Theoretical Astrophysics, 60 St. George Street, Toronto, ON M5S 3H8, Canada}
\affil{Canadian Institute for Advanced Research, 180 Dundas St West, Toronto, ON M5G 1Z8, Canada}
\affil{Dunlap Institute for Astronomy and Astrophysics, University of Toronto, 50 St George Street, Toronto, ON M5S 3H4, Canada}
\affil{Perimeter Institute of Theoretical Physics, 31 Caroline Street North, Waterloo, ON N2L 2Y5, Canada}

\author[0000-0001-9434-3837]{Adam T. Deller}
\affil{Centre for Astrophysics and Supercomputing, Swinburne University of Technology, John St., Hawthorn, VIC S1RR, Australia}
\affil{ARC Centre of Excellence for Gravitational Wave Discovery (OzGrav), Australia}

\correspondingauthor{Rebecca Lin}
\email{lin@astro.utoronto.ca}

\begin{abstract}
We use four observations with the European VLBI network to measure the first precise radio parallax of the Crab Pulsar.
We found two in-beam extragalactic sources just outside the Crab Nebula, with one bright enough to use as a background reference source in our data.
We use the Crab Pulsar's giant pulses to determine fringe and bandpass calibration solutions, which greatly improved the sensitivity and reliability of our images and allowed us to determine precise positional offsets between the pulsar and the background source.
From those offsets, we determine a parallax of $\pi=0.53\pm0.06\rm{\;mas}$ and proper motion of $(\mu_{\alpha},\mu_{\delta})=(-11.34\pm0.06,2.65\pm0.14)\rm{\;mas\;yr^{-1}}$, yielding a distance of $d=1.90^{+0.22}_{-0.18}\rm{\;kpc}$ and transverse velocity of $v_{\perp}=104^{+13}_{-11}\rm{\;km\;s^{-1}}$.
These results are consistent with the \emph{Gaia} 3 measurements, and open up the possibility of far more accurate astrometry with further VLBI observations.

\end{abstract}


\section{Introduction} \label{sec:intro}

The Crab Pulsar (\object{PSR B0531+21}) is one of the youngest pulsars, situated at the heart of the \object{Crab Nebula}, the remnant of supernova SN 1054 \citep{Duyvendak1942, Mayall1942}.
One of the most observed pulsars, it has been continuously monitored by the $13\rm{\;m}$ dish at the Jodrell Bank Observatory since 1984 \citep{Lyne1993}.
The mean radio profile of the pulsar shows multiple components, with the dominant ones being the main pulse (MP) and interpulses (IPs) which are made up of ``giant pulses'', extremely narrow and bright pulses (for a review, see \citealt{Eilek2016}).
The pulse emissions are not only bright in radio but visible up to $\gamma$-ray energies with the MP and IPs showing strong alignment across the full electromagnetic spectrum \citep{Moffett1996}.
The pulsar also undergoes glitches, discrete changes in the pulsar rotation rate, every few years\footnote{\url{http://www.jb.man.ac.uk/pulsar/glitches.html}} (e.g., \citealt{Espinoza2011, Shaw2021}).
This wealth of pulse phenomena offers a great opportunity for understanding the pulsar emission mechanism and possibly constrain the nuclear physics of neutron star interiors.
Additionally, the young age of this system ($\sim\!1000\rm{\;yr}$) makes it the ideal laboratory to study not only the evolution of young pulsars but also pulsar wind nebulae and supernova remnants.

Since the discovery of the Crab Pulsar, there have been several attempts to constrain the distance and proper motion of the pulsar.
For the distance, early attempts based on various lines of evidence including kinematic, spectroscopic and age-related considerations placed the pulsar between $1.4$ and $2.7\rm{\;kpc}$ \citep{Trimble1973}.
From galactic electron density distribution models, the distance to the pulsar can be estimated as $\sim\!1.7\rm{\;kpc}$ with the NE2001 model \citep{Cordes2002} and $\sim\!1.3\rm{\;kpc}$ with the YMW16 model \citep{Yao2017}.

While these estimates give a sense of the distance, none of them are precise and none are direct measurements.
Indeed, many rely on assumptions one would like to test.
For instance, the kinematic constraints implicitly assume a roughly spherical nebula, while the dispersion-measure based distances rely on electron density models.

For the proper motion, similarly early measurements of the Crab Pulsar were relatively poor \citep{Minkowski1970, Wyckoff1977, Caraveo1999}.
A first relatively precise measurement was derived from Hubble Space Telescope observations spanning over a decade, of $(\mu_{\alpha}, \mu_{\delta}) = (-11.8\pm0.4\pm0.5, 4.4\pm0.4\pm0.5)\rm{\;mas\;yr^{-1}}$ \citep{Kaplan2008}.

While a precise parallax and proper motion measurement of the Crab Pulsar would be important, it is impeded by complications in doing astrometry at both radio and optical wavelengths; furthermore, due to the glitches, pulsar timing also cannot help (for a review, see \citealt{Kaplan2008}).
In the optical, this changed with the \emph{Gaia} mission, which presented the first precise astrometry in its second data release (\emph{Gaia} DR2, \citealt{Gaia2018}): $\pi=0.27\pm0.12\rm{\;mas}$ and $(\mu_{\alpha}, \mu_{\delta}) = (-11.8\pm0.2, 2.65\pm0.17)\rm{\;mas\;yr^{-1}}$, respectively.
The precision was improved in the third data release, \emph{Gaia} DR3 \citep{Gaia2022, Antoniadis2020}: $\pi=0.51\pm0.08\rm{\;mas}$ and $(\mu_{\alpha}, \mu_{\delta}) = (-11.51\pm0.10, 2.30\pm0.06)\rm{\;mas\;yr^{-1}}$, respectively.

While impressive, the difference in measured parallax between the two data releases is somewhat worrying.
It might be related to the fact that the measurements are affected by the Crab Pulsar not being a typical optical source, being embedded in an optically bright nebula and producing variable emission near itself that would be only marginally resolved.
Hence, it would be best to have an independent measurement.

At radio wavelengths, Very Long Baseline Interferometry (VLBI) has been very successful in measuring accurate parallaxes and proper motions for pulsars both weaker and further away than the Crab Pulsar (e.g., \citealt{Chatterjee2009, Deller2019}).
For the Crab Pulsar, a difficulty is that it is embedded in a large, $\sim\!6\arcmin\times4\arcmin$, radio-bright nebula.
The high brightness effectively raises the overall system temperature in any observation, making the average emission of the Crab Pulsar hard to detect.
This particularly affects observations at higher frequencies, where the angular resolution is better but the pulse emission fainter ($f_\nu\propto\nu^{-3.1}$, \citealt{Lorimer1995}).
But at lower frequencies, where the pulsar is brighter, the ionosphere hinders astrometry, especially in the absence of an extragalactic source that can be used as an in-beam calibrator -- which has to be outside the nebula, since otherwise it would be severely broadened by scattering.

The problem of a lack of an in-beam calibrator has recently been solved: the Wide-field VLBA Calibrator Survey (WFCS; \citealt{Petrov2021}) lists a suitable nearby source (one which we also discovered independently; see Section~\ref{sec:obs}).
With such an in-beam extragalactic source, one avoids uncertainties in extrapolating phasing solutions for a phase calibrator that is multiple degrees away.
And even if the in-beam calibrator is not very bright, a parallax measurement to within a $0.1\rm{\;mas}$ should be possible if one can self-calibrate on the pulsar \citep{Fomalont1999, Deller2019}.

For the Crab Pulsar, the bright nebula prevents self-calibration on the regular pulse emission (e.g., \citealt{Lobanov2011} used an external phase calibrator for their VLBI imaging).
In principle, the Crab Pulsar's giant pulses can help, as they are extremely bright and can be detected with single dishes.
Because they occur randomly in time, however, even pulsar gating on the corresponding phase windows does not give very good signal-to-noise (S/N) ratios.

In this paper, we present a technique using only the Crab Pulsar's giant pulses to model ionospheric and instrumentation variations for self-calibration, and show that with the newly found nearby extragalactic reference sources this enables precise parallax and proper motion measurements.
In the following, we first describe in Section~\ref{sec:obs} the VLBI data we took, as well as the archival Very Large Array (VLA) dataset we used to search for extragalactic references.
In Section~\ref{sec:data_reduction}, we describe how we correlated our VLBI data to form visibilities, calibrated the visibility data with the giant pulses, and extracted positions of our sources.
In Section~\ref{sec:astrometry}, we derive the parallax and proper motion from the positions.
We compare with the \emph{Gaia} results in Section~\ref{sec:results}, and discuss ramifications and future prospects in Section~\ref{sec:futurework}.

\section{Observations}\label{sec:obs}
\begin{figure*}
  \centering
\includegraphics[width=0.985\textwidth,trim=0 0 0 0,clip]{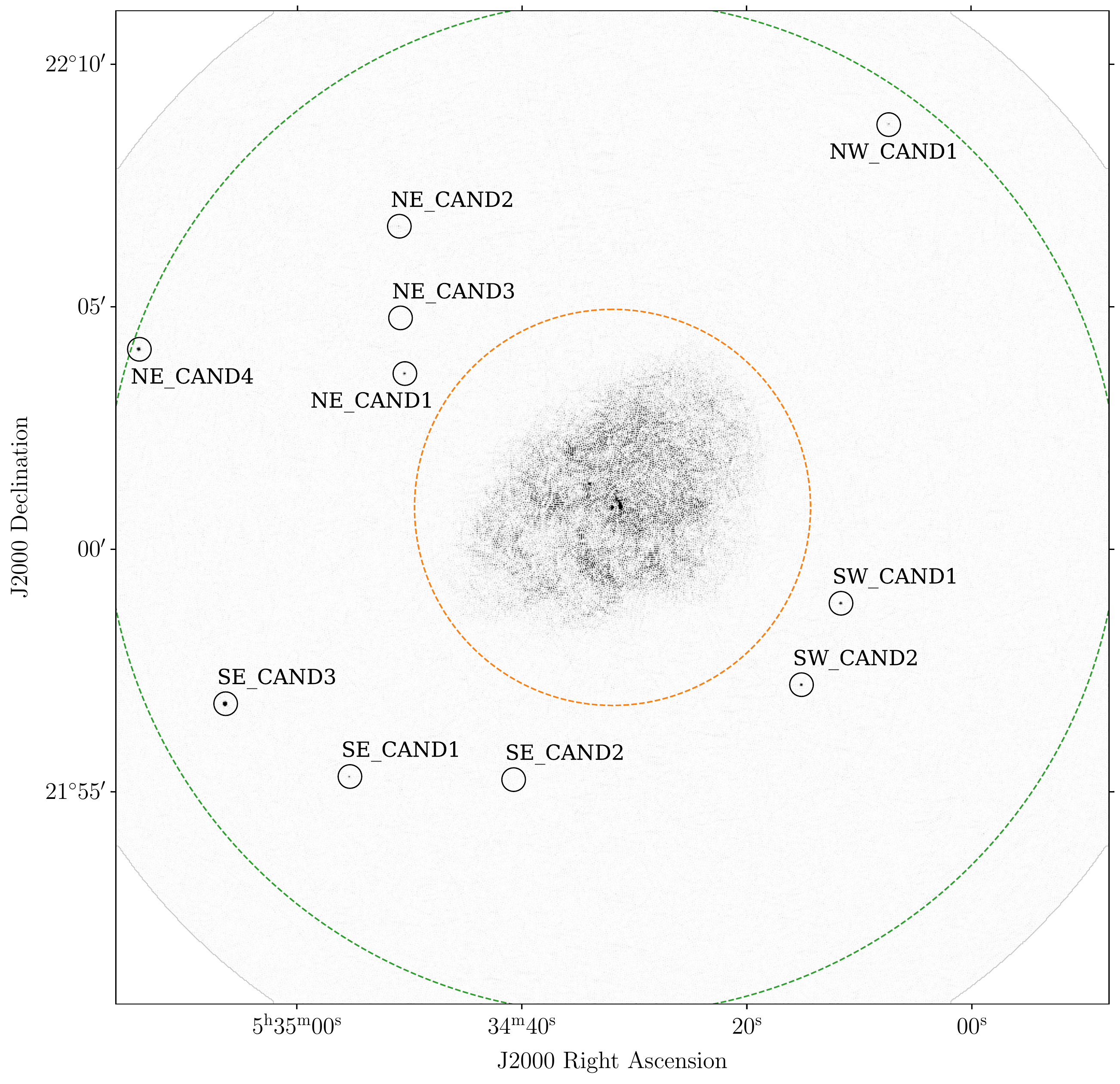}
  \caption{
    VLA image of the Crab Pulsar with our ten candidate reference sources marked (black circles).
    Overlaid are the approximate fields of view of Effelseberg (orange dotted circle) and Badary (green dotted circle), as implied by the full-width-half-maximum (FWHM) at our central observing frequency (using effective apertures of $78\rm{\;m}$ and $30.5\rm{\;m}$, respectively).
    All candidates are outside the field of view of Effelseberg, but within the fields of view of most other dishes, which have diameters comparable to Badary.
    \label{fig:VLA}}
\vspace{3mm}
\end{figure*}

Our observations were taken with the European VLBI Network (EVN) at four epochs between 2015 Oct and 2017 May, using a total of $10\rm{\;hr}$ (see Table~\ref{table:log}).
Real-sampled data in left and right circular polarizations were recorded in either 2-bit MARK 5B or VDIF format at each telescope, except for the $70\rm{\;m}$ at the Robledo Deep Space Station (Ro) where only left circular was available.
The frequency range of $1594.49$--$1722.49\rm{\;MHz}$ was covered, in either eight contiguous $16\rm{\;MHz}$ or four contiguous $32\rm{\;MHz}$ wide bands.

Individual scans on the Crab Pulsar lasted $\sim\!5$ (EK036~C-D) to $\sim\!25\rm{\;min}$ (EK036~A-B), and were interleaved with observations of J0530+1331 ($\sim\!5$ to $\sim\!10\rm{\;min}$, bandpass calibrator source at $8\fdg5$ from the target) and/or J0518+2054 ($\sim\!0.5$ to $\sim\!1\rm{\;min}$, phase calibrator source at $4\fdg0$ from the target).
The unusually long integration times on the target (in particular in EK036~A and B) and short integrations on the phase calibrator were chosen because we only intended the phase calibrator to provide a first crude calibration, just enough to later perform self-calibration on the target; we realized phase calibration at the level required for accurate astrometry would be impossible given the large separation between target and phase calibrator and the bright emission from the Crab Nebula.

After a first inspection of the data from EK036~A, however, it became clear that our initial approach led to phase errors that were too large to perform traditional self-calibration on the Crab Pulsar (see Figure~\ref{fig:dirty_crab_images}).
Hence, we reduced the integration times on the target in the subsequent EK036~C and D observations (e.g., \citealt{Lobanov2011} were able to transfer phase solutions from J0518+2054 using a much shorter calibrator/target cycle of $2/5\rm{\;min}$.).

At the time of these observations, no extragalactic sources near the Crab Pulsar were known that would be suitable as in-beam calibrators.
Therefore, the Crab Pulsar pointings were centered on the pulsar itself (see Table~\ref{table:pointing_centers}), in the hope that suitable in-beam references could be found within the field of view of the smaller participating stations.

\begin{deluxetable*}{lllclccrr}
\tabletypesize{\small}
\setlength{\tabcolsep}{3.2pt}
\tablecaption{Observation and Giant Pulse Log\label{table:log}}
\tablenum{1}
\tablehead{\colhead{Observation}&
\colhead{Date}&
\colhead{MJD}&
\colhead{$t_{\text{exp}}$\tablenotemark{a}}&
\colhead{$t_{\text{target}}$\tablenotemark{b}}&
\colhead{Telescopes used\tablenotemark{c}}&
\colhead{DM\tablenotemark{d}}&
\multicolumn{2}{c}{Giant Pulses\tablenotemark{e}}\\[-.7em]
\colhead{code}&
&
&
\colhead{(h)}&
\colhead{(h)}&
&
\colhead{($\rm{pc\;cm^{-3}}$)}&
\colhead{$N$}&
\colhead{$r$ ($\rm{min^{-1}}$)}}
%% All data must appear between the \startdata and \enddata commands
\startdata
EK036~A & 2015 Oct 18 & 57313.96 & 4 & 3.27 & Ef\hx{3}Bd\hx{3}Hh\hx{3}Jb\hx{11}Mc\hx{3}O8\hx{3}Ro*\hx{3}Sv\hx{3}T6*\hx{3}Tr\hx{16}Wb\hx{3}Zc & 56.7772 & \hspace{0.5em}  686 &  3.50 \\
EK036~B & 2016 Oct 31 & 57692.98 & 2 & 1.65 & Ef\hx{3}Bd\hx{3}Hh\hx{22}Mc\hx{3}O8\hx{21}Sv\hx{45}Wb\hx{3}Zc                                  & 56.7668 & \hspace{0.5em} 1067 & 10.81 \\
EK036~C & 2017 Feb 25 & 57809.67 & 2 & 1.15 & Ef\hx{3}Bd\hx{3}Hh\hx{3}Jb\hx{12}Mc\hx{3}O8\hx{21}Sv\hx{32}Ur\hx{3}Wb\hx{3}Zc                  & 56.7725 & \hspace{0.5em}  281 &  4.08 \\
EK036~D & 2017 May 28 & 57901.40 & 2 & 1.24 & Ef\hx{3}Bd\hx{3}Hh\hx{3}Jb-II\hx{3}Mc\hx{3}O8\hx{21}Sv\hx{21}Tr\hx{3}Ur\hx{3}Wb\hx{3}Zc        & 56.7851 & \hspace{0.5em}  740 &  9.94
\enddata
\tablenotetext{a}{Total observing time, including telescope setup and calibration.}
\tablenotetext{b}{Total exposure on target.}
\tablenotetext{c}{
  We omit telescopes where data were corrupt, where significant RFI occurred and/or where we were unable to determine reliable fringe solutions.
  Asterisks beside a telescope indicate that the telescope was unable to see the source for the full observing time; furthermore, Ro had left-circular polarization only.
  Abbreviations are: Ef: the~$100\rm{\;m}$ Effelsberg telescope; Bd: the~$32\rm{\;m}$ at Badary; Hh: the~$26\rm{\;m}$ in Hartebeesthoek; Jb: the~$76\rm{\;m}$ Lovell telescope; Jb-II: the~$25\rm{\;m}$ Mark II Telescope at the Jodrell Bank Observatory; Mc: the~$32\rm{\;m}$ at Medicina; O8: the~$25\rm{\;m}$ at Onsala; Ro: the~$70\rm{\;m}$ at the Robledo Deep Space Station; Sv: the~$32\rm{\;m}$ at Svetloe; T6:~$65\rm{\;m}$ at Tianma; Tr: the~$32\rm{\;m}$ at Toru\'n; Wb: the~$25\rm{\;m}$ RT1 telescope at Westerbork; and Zc: the~$32\rm{\;m}$ at Zelenchukskaya.
}
\tablenotetext{d}{Inferred from the giant pulses.}
\tablenotetext{e}{Total number and rate of giant pulses (including both MP and IP) found using a detection threshold of $50\sigma$ on incoherently summed data (for details, see \citealt{Lin2023}).}
\vspace{3.8mm}
\end{deluxetable*}

\begin{deluxetable}{cllc}
\tabletypesize{\small}
\setlength{\tabcolsep}{2pt}
\tablecaption{Target and Calibrator Scan Pointing Centers\label{table:pointing_centers}}
\tablenum{2}
\tablehead{\colhead{Source}&
  \colhead{Right Ascension}&
  \colhead{Declination}&
  \colhead{Sep.}\\[-.7em]
  &
  \colhead{($\alpha$)}&
  \colhead{($\delta$)}&
  \colhead{(\arcdeg)}}
%% All data must appear between the \startdata and \enddata commands
\startdata
PSR B0531+21 & $05^{\rm{h}}34^{\rm{m}}31.934^{\rm{s}}$     & 22\arcdeg00\arcmin52.191\arcsec    & \nodata\\
J0530+1331   & $05^{\rm{h}}30^{\rm{m}}56.4167465^{\rm{s}}$ & 13\arcdeg31\arcmin55.149516\arcsec & 8.5\\
J0518+2054   & $05^{\rm{h}}18^{\rm{m}}03.8245128^{\rm{s}}$ & 20\arcdeg54\arcmin52.497365\arcsec & 4.0\\
\enddata
\tablecomments{Coordinates listed here are in the J2000 FK5 frame. The separation between the calibrator sources and the Crab Pulsar is given in the last column.}
\end{deluxetable}

\begin{deluxetable}{lllrr}
\tabletypesize{\small}
\setlength{\tabcolsep}{3.75pt}
\tablecaption{Candidate Extragalactic Sources from VLA Data\label{table:VLA_sources}}
\tablenum{3}
\tablehead{\colhead{Source}&
  \colhead{Right Ascension}&
  \colhead{Declination}&
  \colhead{Peak}&
  \colhead{Sep.}\\[-.7em]
  &
  \colhead{($\alpha$)}&
  \colhead{($\delta$)}&
  \colhead{S/N\tablenotemark{a}}&
  \colhead{(\arcmin)}}
%% All data must appear between the \startdata and \enddata commands
\startdata
NE\_CAND1 & $05^{\rm{h}}34^{\rm{m}}50.43^{\rm{s}}$ & 22\arcdeg03\arcmin37.58\arcsec &  16.3 &  5.1\\
NE\_CAND2 & $05^{\rm{h}}34^{\rm{m}}50.93^{\rm{s}}$ & 22\arcdeg06\arcmin39.79\arcsec &   8.4 &  7.3\\
NE\_CAND3 & $05^{\rm{h}}34^{\rm{m}}50.80^{\rm{s}}$ & 22\arcdeg04\arcmin46.41\arcsec &   4.9 &  5.9\\
NE\_CAND4 & $05^{\rm{h}}35^{\rm{m}}14.05^{\rm{s}}$ & 22\arcdeg04\arcmin07.43\arcsec &  42.6 & 10.3\\
SE\_CAND1 & $05^{\rm{h}}34^{\rm{m}}55.31^{\rm{s}}$ & 21\arcdeg55\arcmin18.94\arcsec &  14.8 &  7.8\\
SE\_CAND2 & $05^{\rm{h}}34^{\rm{m}}40.74^{\rm{s}}$ & 21\arcdeg55\arcmin15.31\arcsec &   6.5 &  6.0\\
SE\_CAND3 & $05^{\rm{h}}35^{\rm{m}}06.34^{\rm{s}}$ & 21\arcdeg56\arcmin49.11\arcsec & 463.0 &  8.9\\
NW\_CAND1 & $05^{\rm{h}}34^{\rm{m}}07.34^{\rm{s}}$ & 22\arcdeg08\arcmin45.63\arcsec &  19.8 &  9.7\\
SW\_CAND1 & $05^{\rm{h}}34^{\rm{m}}11.62^{\rm{s}}$ & 21\arcdeg58\arcmin53.38\arcsec &  18.7 &  5.1\\
SW\_CAND2 & $05^{\rm{h}}34^{\rm{m}}15.14^{\rm{s}}$ & 21\arcdeg57\arcmin12.71\arcsec &  14.0 &  5.3
\enddata
\tablecomments{Coordinates listed here are in the J2000 FK5 frame. SE\_CAND3 and NE\_CAND4 where later confirmed to be visible in the EVN data. The separation between the candidate sources and the Crab Pulsar is given in the last column.}
\tablenotetext{a}{As found in the VLA data.}
\end{deluxetable}

\vspace{-17mm}
Given the high resolution of the EVN data, an untargeted search for in-beam sources is nearly intractable.
Instead, we searched for candidates in an archival VLA dataset (project code 12B-380), taken in A-array configuration on 2012 Nov.\ 26 \& 27 at $3\rm{\;GHz}$ (S band, covering 2--4~GHz).
We used the standard Common Astronomy Software Applications VLA calibration pipeline (\casa\ 5.1.1, \citealt{CASA2022}) to perform automatic flagging and calibration of the two datasets.
No careful flux calibration was applied.
After inspection of the data, we decided to focus only on the later run, from 2012~Nov.~27.
We used the \casa\ task \texttt{tclean} for imaging, limiting ourselves to the lower half of the frequency band, i.e., 2--3$\rm{\;GHz}$.
Moreover, we limited the $uv$ range, excluding visibilities from baselines $<\!75\rm{\;k\lambda}$ in order to filter out the extended emission from the Crab Nebula itself.
Since our aim was to find compact sources within the field of view of the VLA, we generate an image of $8192\times8192$ pixels at an angular resolution of $0.15\rm{\;arcsec/pixel}$, oversampling by about a factor 3 the $477\times470\rm{\;mas}$ beam (position angle $-60\arcdeg$).
The rms in the final image varies by a factor of up to 4 between the central region and the outer region of the image because of the Crab Nebula's emission.

We exported the cleaned image as a FITS file and searched for radio sources by normalizing the image relative to a median-filtered version to pick out outliers.
In this way, we found ten candidates, which we list in Table~\ref{table:VLA_sources} and show in Figure~\ref{fig:VLA}.

For all of the candidates, we created images from our EVN data, finding that the two brightest ones were detected: SE\_CAND3 and NE\_CAND4 (see Section~\ref{subsec:imaging} and Figure~\ref{fig:clean_images}).
We were able to find a source at the location of SE\_CAND3 in the new Wide Field Very Long Baseline Array (VLBA) calibrator survey \citep{Petrov2021}, which lists it as the compact source WFCS~J0535+2156, and in the Very Large Array Sky Survey (VLASS) \citep{Gordon2021} as VLASS1QLCIR J053506.32+215649.3.
Candidate NE\_CAND4 was also seen in VLASS as VLASS1QLCIR J053514.04+220407.7, but not in the VLBA catalogue.

Looking through other catalogues, we found sources matching the position of SE\_CAND3 in the Wide-field Infrared Survey Explorer Data Release \citep{Cutri2012} and in the UKIRT Infrared Deep Sky Survey \citep{Lawrence2012}.
It also has a counterpart in \emph{Gaia} DR3 \citep{Gaia2022}, with a parallax and proper motion consistent with zero.
Thus, it seems likely that SE\_CAND3 is an active galactic nucleus.

\section{Correlation, Calibration, Images, Positions and Uncertainties}\label{sec:data_reduction}

\subsection{Visibilities}\label{subsec:SFXC}

We correlated the data from the different telescopes using the publicly available Super FX Correlator (\sfxc\ 5.1; \citealt{Keimpema2015}), which prior to preforming the correlations, corrects for station clock offsets and rates, as well as for geometric delays using CALC10\footnote{\url{https://space-geodesy.nasa.gov/techniques/tools/calc_solve/calc_solve.html}} \citep{Ryan1980}.
At this stage, no additional station-specific delays or atmospheric distortions of the wavefront are taken into account.

For each observation, two correlation passes were performed.
The first pass correlated on the Crab Pulsar in pulsar gating mode (described below), and on the bandpass and phase calibrators in ungated mode.
The correlation centers in this pass are the same as the antenna pointing centers given in Table~\ref{table:pointing_centers}.
In the second correlator pass, we correlated all target scans again, but now ungated and centred on the locations of our candidate sources (see Table~\ref{table:VLA_sources}), using \sfxc's multi-phase center mode.

For the pulsar gating, we created polyco files using {\sc Tempo2} \citep{Hobbs2012} with the Crab Pulsar ephemeris, starting from the ephemeris provided by Jodrell Bank Observatory\footnote{\url{http://www.jb.man.ac.uk/~pulsar/crab.html}} \citep{Lyne1993} and then adjusting the phase and dispersion measure to values found in \cite{Lin2023} for the same data (see Table~\ref{table:log}).
With these, we used \sfxc\ to incoherently de-disperse\footnote{\sfxc\ version 5.1 does not have coherent de-dispersion capabilities.}, fold, and gate the pulsar observations on the MP phase window ($2.1\%$ of the $\sim\!33\rm{\;ms}$ pulse period).

With the gated mode, we gain S/N by removing time ranges when little if any pulsar signal is present.
However, since giant pulses are short in duration (most of the signal is within the scattering timescale of $\sim\!5\rm{\;\mu s}$ at our observing frequency, see \cite{Lin2023} for examples of giant pulses from these datasets) and occur only in some pulse rotations, one could in principle get much better S/N ratio by only including pulse rotations in which giant pulses occur.
Furthermore, one could also include IP giant pulses and possibly other pulse components.
We did not pursue these potential improvements, since we find below (in Section~\ref{subsec:imaging}) that the S/N ratio of the images created from the MP gated visibilities is much larger than that of the in-beam candidate sources, and thus does not limit the accuracy of the astrometry.

For all correlations, we used a spectral resolution of $4096$ channels across the total bandwidth, limiting any dispersive in-channel smearing to $3\rm{\;\mu s}$ (of order a giant pulse width).
We used a temporal resolution of $0.5\rm{\;s}$ to have the option of, in post-processing, select only time integrations where particularly bright giant pulses occurred (but we did not use this, as the S/N ratio sufficed).
In total, for each observation eleven visibility sets were created, one for the Crab Pulsar and one each for our candidate sources.
Calibrator visibility data were included in each visibility set, hence each set contained three sources.

\subsection{Calibration}\label{subsec:calibration}

We calibrated our visibilities with the help of \casa\ 6.5, writing custom calibration scripts to ensure that our calibrations are consistent across all observations and to help track our configurations.

In preparation, we first converted visibility data to \casa\ Measurement Sets using Joint Institute for VLBI in Europe (JIVE) post-processing tools, and set up antenna tables with diameters and axis offsets from the station summary files.
We also set up amplitude calibration tables with system temperature, gain curve and primary beam corrections.
Since system temperature and gain curve measurements from the telescope logs were affected by the bright Crab Nebula and thus unreliable (and some were simply missing), we instead used nominal values taken from the EVN status table\footnote{\url{http://old.evlbi.org/user_guide/EVNstatus.txt}} and included the flux density of the Crab Nebula $S_{\text{CN}}=955\nu^{-0.27}\rm{\;Jy}$, where $\nu$ is our observing frequency in $\rm{GHz}$ \citep{Bietenholz1997}.
As our goal is precise astrometry, the true flux density of our sources is of little importance and this flux scaling is sufficient for estimating which of the candidate sources will likely be visible in the EVN datasets.
We flagged times and frequencies where the signal was poor (i.e., before the start and end of each scan, and at passband edges), as well as particularly strong radio frequency interference (RFI) previously detected in the baseband data (see \citealt{Lin2023}), taking care to ensure that giant pulse signals were not accidentally removed.
Finally, to account for the reduced sensitivity away from the antenna pointings, we applied a primary beam correction for our in-beam candidate correlation centers (assuming an Airy disk with effective aperture sizes provided by the JIVE team, separately for each of our eight spectral windows).

For calibration, we started by determining phase and delay corrections due to instrument and atmospheric variations towards our calibrator sources: we use \casa's \texttt{fringefit} task to determine solutions in $60\rm{\;s}$ intervals for each spectral window and polarization independently, with Effelsberg as the reference antenna.
We attempted transferring the fringe solutions to the Crab Pulsar, but found relatively poor results (see Figure~\ref{fig:dirty_crab_images} and Section~\ref{subsec:imaging} below).
This was not unexpected given that our calibrators are far from the Crab Pulsar and that the target scans are relatively long compared to the timescale of a few minutes of ionospheric variations.

\begin{figure}
\centering
\includegraphics[width=0.47\textwidth,trim=0 0 0 0,clip]{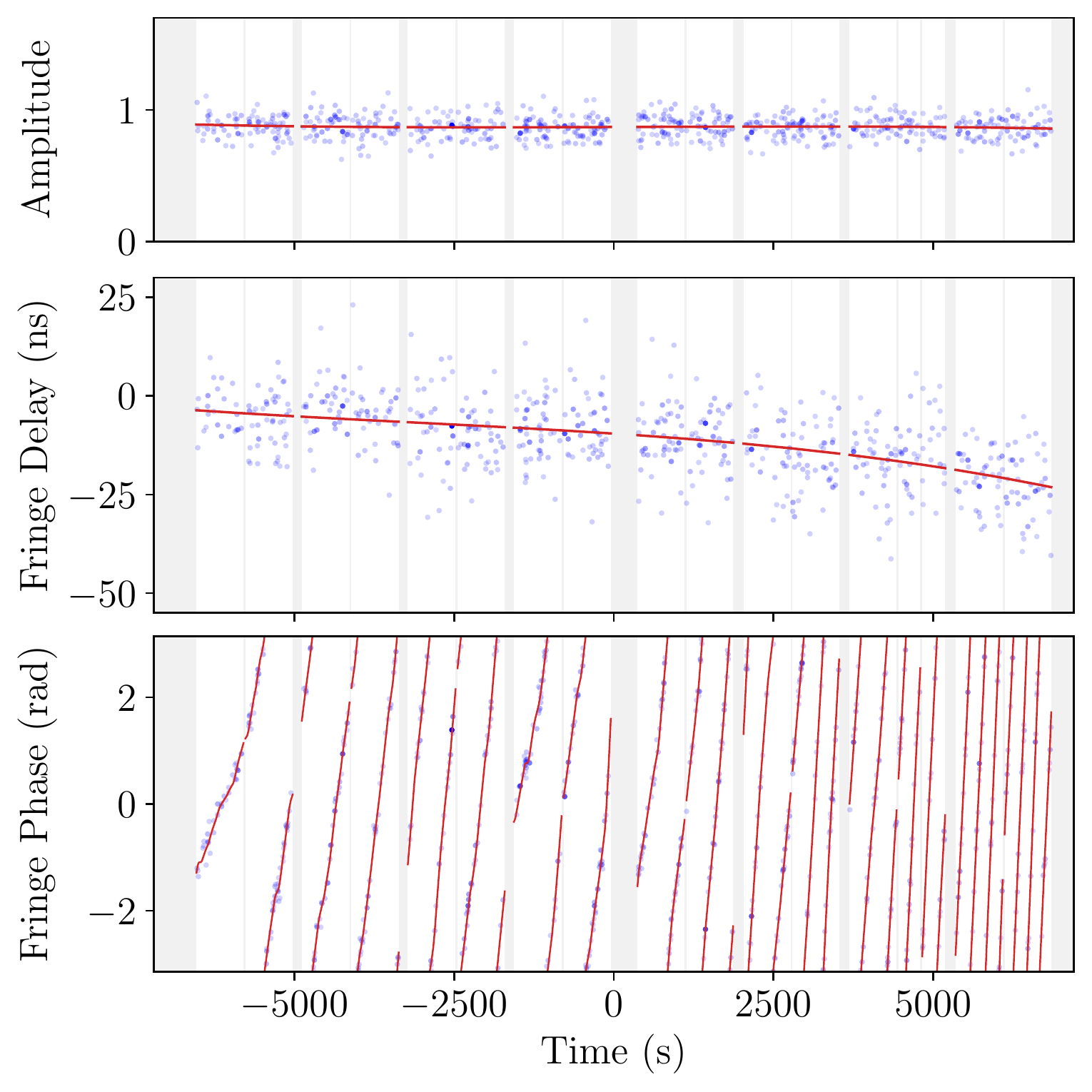}
\caption{
 Amplitude (top), fringe delay (middle) and fringe phase (bottom) of Badary relative to Effelsberg derived from giant pulses of EK036~A for the frequency band $1626.49\!-\!1642.49\rm{\;MHz}$ in left circular polarization (blue points).
 The opacity of the individual blue points scales with the square root of the S/N of the giant pulse.
 The red lines shows our fits.
 The gray shaded regions indicate when the telescope was not observing the Crab Pulsar.
 The fringe rate increases strongly near the end as the Crab Pulsar is setting at Badary.\label{fig:fringe_solutions}}
\end{figure}

\begin{figure}
\centering
\includegraphics[width=0.47\textwidth,trim=0 0 0 0,clip]{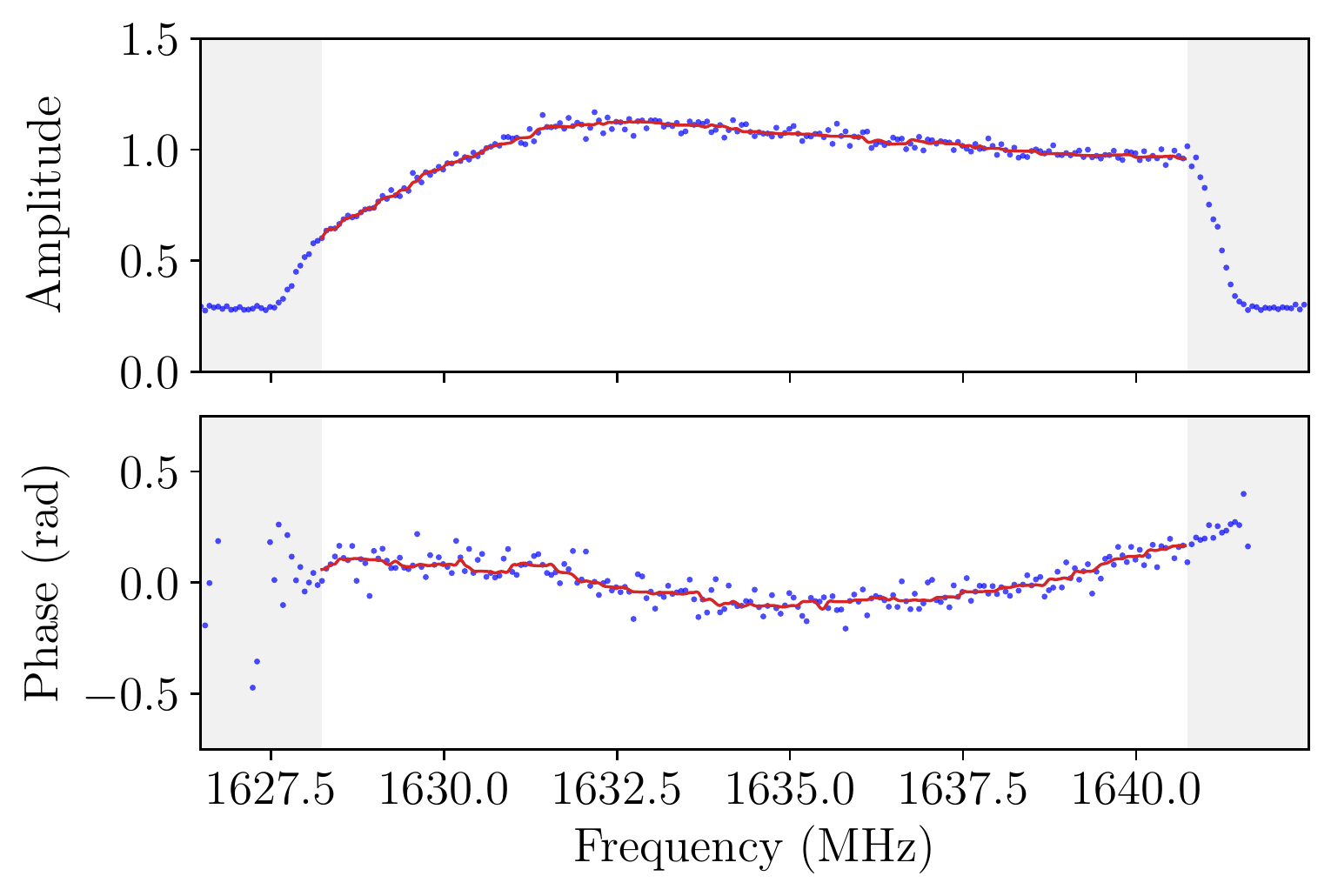}
\caption{
  Complex bandpass amplitude (top) and phase (bottom) of Lovell relative to Effelsberg derived from giant pulses of EK036~A for the frequency band $1626.49\!-\!1642.49\rm{\;MHz}$ in left circular polarization (blue points).
  The red lines shows our fits.
  The gray shaded regions indicate were we flagged data in \casa\ as little signal is detected and the passband rolls off.\label{fig:bandpass_solutions}}
\end{figure}

\begin{figure*}[ht]
  \centering
  \includegraphics[width=0.985\textwidth,trim=0 0 0 0,clip]{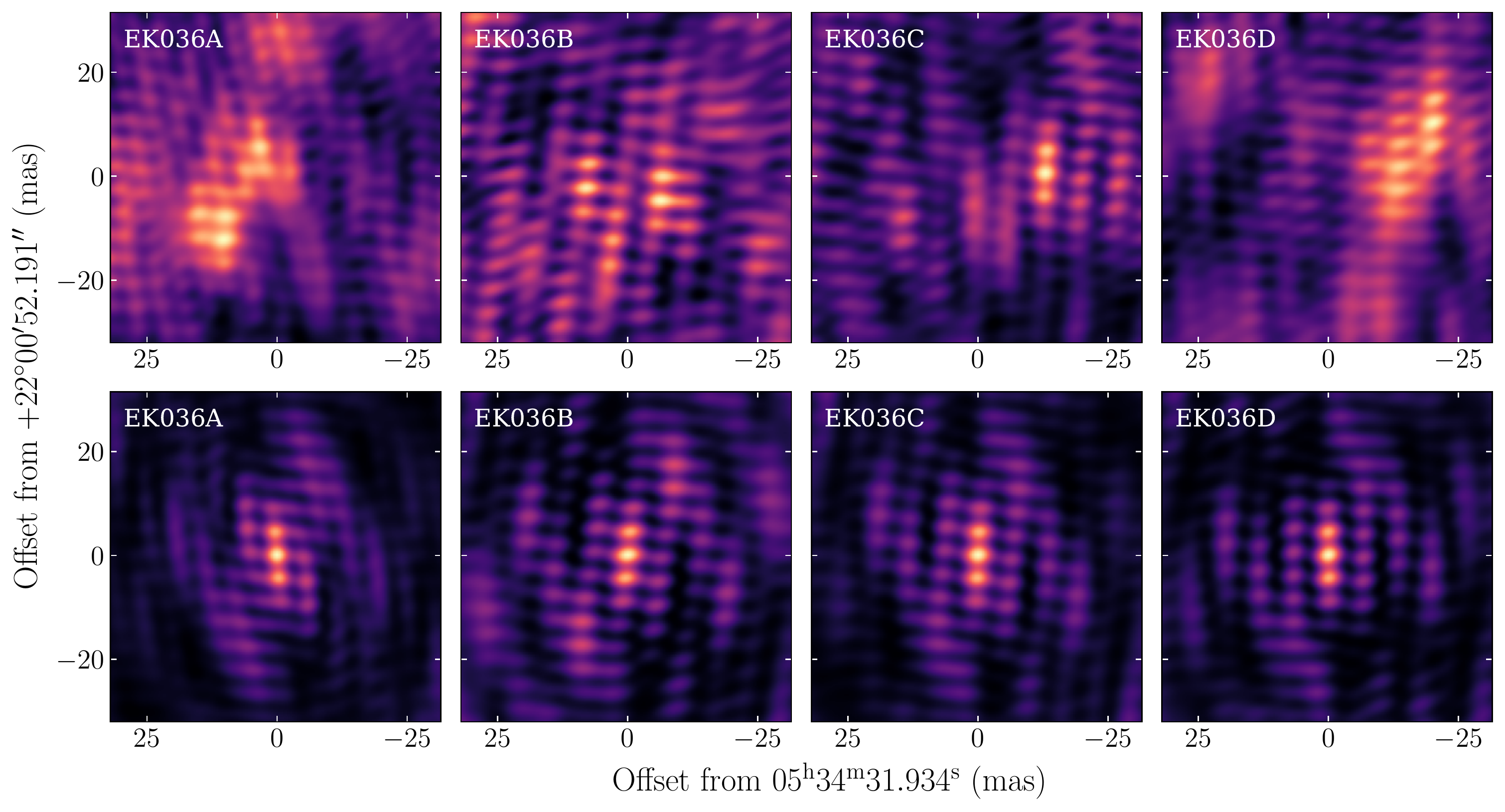}
  \caption{
    Dirty images of the Crab Pulsar for our four observations, comparing calibration solutions transferred from calibrator sources with those using giant pulses for self-calibration.
    {\em Top:\/} In general, transfer of the calibrator solutions to the Crab Pulsar resulted in poor images, particularly for EK036~A and EK036~B where the calibrator/target cycle was $\sim\!0.5/25\rm{\;min}$ and ionospheric variations could not be modeled well.
    In EK036~C and EK036~D, the calibrator/target cycle was $\sim\!1/5\rm{\;min}$ resulting in somewhat better dirty images.
    {\em Bottom:\/} Using calibration solutions derived directly from the Crab Pulsar's giant pulses significantly improves the dirty images, with the Crab Pulsar clearly located in the center.
    \label{fig:dirty_crab_images}
  }
\vspace{3mm}
\end{figure*}

Since the Crab Pulsar is the brightest of the in-beam sources, we instead used it to self-calibrate.
We first tried using the gated pulsar visibilities from \sfxc, but these do not have sufficient S/N on short integrations and hence the resulting fringe solutions obtained using \casa's \texttt{fringefit} task were unreliable, showing extreme variations without any discernible pattern.

Instead, we follow \cite{Lin2023} and use giant pulses to model the delays, amplitude and phase rotations in each spectral window, and write these models to \casa\ compatible fringe and amplitude tables.
Specifically, we use all giant pulses (both MP and IP) detected with a S/N ratio of 50 in the incoherently summed telescope data (a cut-off that ensures no false detections; we find no pulses outside of the MP and IP phase windows, see \citealt{Lin2023} for details on the data reduction giant and giant pulse detection).
We show an example of an extreme fringe solution in Figure~\ref{fig:fringe_solutions} (for Badary in the EK036~A observation, where the Crab Pulsar is setting, causing a rapid increase in path length through the ionosphere thus a large increase in fringe rate).
We applied these solutions to both the Crab Pulsar and the candidate in-beam source visibility sets.

Since our detection rate is high, at $\sim\!4\!-\!11$ every minute (see Table~\ref{table:log}), we can easily follow ionospheric variations towards the Crab Pulsar and thus, unlike many calibration pipelines, do not apply archival global ionosphere models such as the ionosphere vertical total electron content (TEC) maps from NASA's Crustal Dynamic Data Information System (CDDIS)\footnote{\url{https://cddis.nasa.gov/Data_and_Derived_Products/GNSS/atmospheric_products.html\#iono}} to the pulsar.
We thus avoid uncertainties associated with the coarse resolution of the TEC maps (5\arcdeg\ in longitude by 2.5\arcdeg\ in latitude and $2\rm{\;hr}$ temporal resolution), the accuracy of $\sim\!2$ to $8\rm{\;TECU}$, and modeling assumptions required in using it (in \casa, that the ionosphere is a thin shell at a constant height of $450\rm{\;km}$).
Indeed, unreliable TEC information can sometimes produce negative parallaxes with smaller errors (e.g. \citealt{Deller2009}), and recent analysis by \cite{Petrov2023} found that while TEC maps certainly can help improve measurements, corrections to the default values and implementation were needed to obtain the best absolute astrometry.

Still, even for our close in-beam reference source, the ionospheric contributions will differ slightly between it and the Crab Pulsar.
Thus, while for our main analysis described below we do not use the TEC maps, we describe a separate analysis applying a relative correction based on them in Appendix~\ref{sec:appendix}.
We find that this gives consistent astrometric results with those of Section~\ref{sec:astrometry} below.

For bandpass calibration (i.e., time-independent frequency calibration), we again use our giant pulses, this time creating visibilities (for details, see \citealt{Lin2023}), which we then averaged across time to solve for the complex bandpass.
As before, we use Effelsberg as our reference antenna.
The complex bandpass was smoothed with a median filter to remove any remaining RFI contributions, then modeled with a simple cubic spline interpolation and normalized to unity to preserve the flux density scale.
We wrote our solutions to \casa\ compatible bandpass tables and applied the corrections to both the Crab Pulsar and the candidate in-beam source visibility sets generated by \sfxc.
We show a typical bandpass solution in Figure~\ref{fig:bandpass_solutions}.

To verify our solutions, we also determined bandpass solutions using \casa's \texttt{bandpass} task on the calibrator sources (using Effelsberg as the reference antenna).
We found that there were no significant differences between these solutions and the ones determined from the giant pulses.
Since we do not use the calibrators elsewhere in our analysis, we decided to stick with the giant-pulse bandpass solutions.

Lastly, we used \casa's \texttt{gaincal} task with a solution interval of $\sim\!5\rm{\;mins}$ to refine our amplitudes on the gated Crab Pulsar visibilities.
Overall, we find that this final amplitude correction showed no variations related to scintillation as in \cite{Deller2009}, as expected given that the Crab Pulsar's scintillation decorrelation bandwidth is much smaller that the width of our subbands.
As before, this adjustment to the absolute flux density scaling should not affect positions (which we confirmed by omitting this step), but does improve the S/N of images slightly, by $\lesssim5\%$.
We again apply these solutions to both the Crab Pulsar and candidate reference sources.

After calibration, we reduced the data to a more manageable size by lowering the spectral resolution to $1024$ channels and the temporal resolution to $2\rm{\;s}$.

\subsection{Images}\label{subsec:imaging}

\begin{figure*}
  \centering
  \includegraphics[width=0.985\textwidth,trim=0 0 0 0,clip]{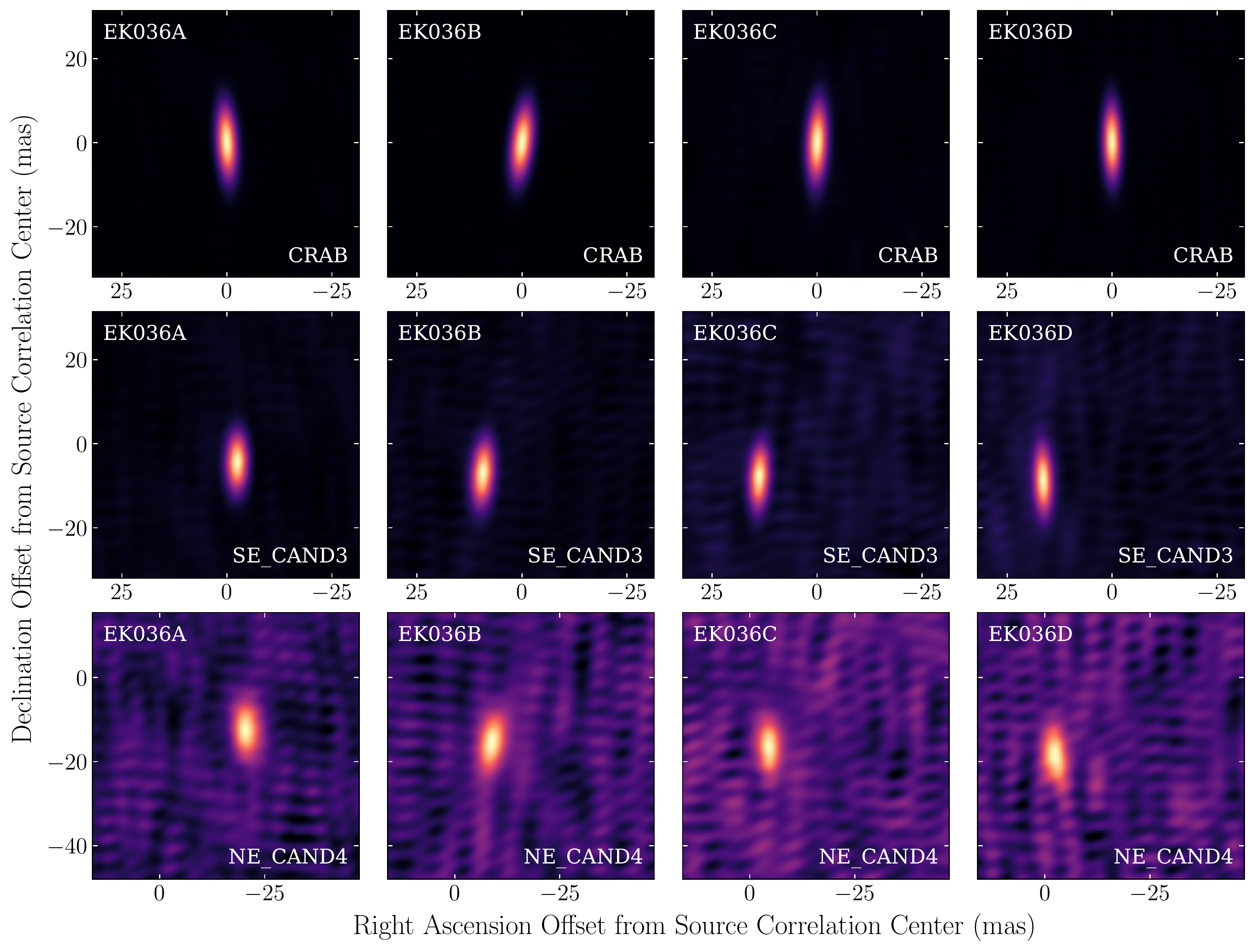}
  \caption{
    Clean images of the Crab Pulsar and candidate reference sources SE\_CAND3 and NE\_CAND4.
    The correlation centers for the pulsar are at the origin by construction.
    The centers for the other two are relative to those of the pulsar and thus these sources show reflex motion (for details, see Section~\ref{subsec:SFXC}).
    \label{fig:clean_images}
  }
\vspace{3mm}
\end{figure*}

All imaging was done using \casa's \texttt{tclean} task.
For all sources, we started with our full bandwidth, and used natural weighting to optimize S/N.
The synthesized beam is similar in all observation, with full width at half maximum of roughly $4\rm{\;mas}\times12\rm{\;mas}$, elongated in declination.
To adequately sample this beam, we use a pixel size of $0.5\rm{\;mas}$ for our images.
All our generated images are $4096\times4096$ pixels in size.

We first formed dirty images for all our visibility sets.
For the Crab Pulsar, after applying the initial calibrations to the visibilities, we applied further calibrations in two separate ways: one using only solutions inferred from the calibrator sources (i.e., phase-referencing), and one using the solutions obtained from giant pulses (i.e., self-calibration).
We compare the resulting dirty images of the Crab Pulsar in Figure~\ref{fig:dirty_crab_images}.
One sees that our giant pulse self-calibration provides much better results.
Thus, for the candidate reference sources, after applying the initial calibrations, we applied further calibrations using only solutions obtained from giant pulses (i.e., effectively phase-referenced relative to the Crab Pulsar).

From these dirty images, we were only able to confidently see two of the in-beam candidate sources, SE\_CAND3 and NE\_CAND4.
This is perhaps unsurprising, since the other candidate sources are much weaker (see Table~\ref{table:VLA_sources}) and our average sensitivity limit is quite poor: even away from the nebula, the rms is $\sim\!0.25\rm{\;mJy/beam}$.
Another possibility is that some of these sources are extended beyond our largest angular scales ($\sim\!140\rm{\;mas}$) and hence resolved out.

For SE\_CAND3, we measured fluxes between $\sim\!13$ and $24\rm{\;mJy}$ in our four epochs, while for NE\_CAND4, we found fluxes between $\sim\!3$ and $5\rm{\;mJy}$,
For comparison, for SE\_CAND3, \cite{Petrov2021} gives $4.3$ and $7.6\rm{\;GHz}$ fluxes of $\sim\!42$ and $\sim\!36\rm{\;mJy}$, respectively, in the WFCS, while \citet{Gordon2021} finds $3\rm{\;GHz}$ fluxes of $\sim\!39$ and $\sim\!4\rm{\;mJy}$ in VLASS for SE\_CAND3 and NE\_CAND4, respectively.
These fluxes seem roughly consistent, taking into account our approximate flux calibration, differences in observing frequency and resolution, as well as possible source variability and structure.

The positions of SE\_CAND3 and NE\_CAND4 are within $50\rm{\;mas}$ of their correlation centers (see Figure~\ref{fig:clean_images}) and well within the uncertainties of positions measured in the VLA data.
Thus, phase drifts resulting from the sources not being exactly at their correlation center are negligible \citep{Fomalont1999} and we do not re-correlate any data.

\begin{deluxetable*}{ccccc}[ht]
\tabletypesize{\small}
\setlength{\tabcolsep}{5pt}
\tablecaption{Relative Positions between the Reference Sources and the Crab Pulsar\label{table:relative_position_fits}}
\tablenum{4}
\tablehead{\colhead{Observation}&
\multicolumn{2}{c}{\dotfill SE\_CAND3 \dotfill}&
\multicolumn{2}{c}{\dotfill NE\_CAND4 \dotfill}\\[-.7em]
\colhead{code}&
\colhead{$\Delta\alpha^{\ast}$ ($\rm{mas}$)}&
\colhead{$\Delta\delta$ ($\rm{mas}$)}&
\colhead{$\Delta\alpha^{\ast}$ ($\rm{mas}$)}&
\colhead{$\Delta\delta$ ($\rm{mas}$)}}
%% All data must appear between the \startdata and \enddata commands
\startdata
EK036~A & $-478472.866\pm0.013\pm0.00\pm0.04$            & $243085.54\pm0.04\pm0.07\pm0.14$ & $-585690.43\pm0.13$           & $-195268.6\pm0.3$\\
EK036~B & $-478484.651\pm0.02\phantom{0}\pm0.02\pm0.04$  & $243088.25\pm0.06\pm0.14\pm0.14$ & $-585702.27\pm0.2\phantom{0}$ & $-195265.4\pm0.5$\\
EK036~C & $-478489.143\pm0.03\phantom{0}\pm0.03\pm0.04$  & $243089.04\pm0.09\pm0.11\pm0.14$ & $-585706.66\pm0.18$           & $-195264.3\pm0.4$\\
EK036~D & $-478491.763\pm0.02\phantom{0}\pm0.03\pm0.04$  & $243089.84\pm0.09\pm0.09\pm0.14$ & $-585708.76\pm0.19$           & $-195262.1\pm0.5$\\
\enddata
\tablecomments{
  All right ascension offset are calculated at the declination of the pulsar.
  For SE\_CAND3, we provide the measurement errors inferred from the fits to the cleaned images, and estimates of the intra-epoch (see Section~\ref{subsec:positions}) and inter-epoch errors (see Section~\ref{sec:astrometry}) , respectively.
  These should be added in quadrature to obtain the total uncertainty.
  For NE\_CAND4, we list only the errors from the position fit, since they are substantially larger than any systematic effects.}
\vspace{3mm}
\end{deluxetable*}

To clean our images of the Crab Pulsar, SE\_CAND3 and NE\_CAND4, we apply a single elliptical mask the size and orientation of the synthesized beam centered on the peak flux in the dirty images to guide the cleaning.
The cleaning was stopped when the residual reached an rms equal to that of a $4096\times4096$ pixel dirty map centered $\sim\!2\arcsec$ West from the source (this is far enough away that there are no sources in the map and side-lobe effects do not affect the field significantly so the average rms measurement is more accurate; the noise level was measured using \casa's \texttt{imstat} task).

We show our clean images of the Crab Pulsar and the two in-beam candidates SE\_CAND3 and NE\_CAND4 in Figure~\ref{fig:clean_images}.
Since we self-calibrated on the pulsar, its position is fixed to the antenna pointing position (see Table~\ref{table:pointing_centers}).
Assuming an extragalactic origin of the in-beam candidate sources, one expects them to move slightly between epochs.
As can be seen in Figure~\ref{fig:clean_images}, this is indeed the case.

\subsection{Positions and their Uncertainties}\label{subsec:positions}

We first tried fitting the cleaned source images with elliptical Gaussians using the \casa\ task \texttt{imfit} which is based on the procedure of \cite{Condon1997}.
However, we found that the position errors provided by \texttt{imfit} were odd -- we expected errors in right ascension and declination to scale with their respective beam sizes, but found that the ratio was substantially different (with errors in declination a factor 7-10 times larger than those in right ascension, instead of the expected factor of $\sim\!3$).
We compared \casa's \texttt{imfit} results with those from the \texttt{jmfit} task from the Astronomical Image Processing System (\aips; \citealt{AIPS1999}) which is also based on \cite{Condon1997}.
The fitted positions were consistent, but the uncertainties from \aips's \texttt{jmfit} do have the expected scaling with beam size.

To investigate this discrepancy in position uncertainties, we implemented our own elliptical Gaussian fit routine in {\sc python}.
We discovered that the discrepancy between \casa\ and \aips\ comes from how the noise and restoring beams are used when determining S/N.
We concluded that for a point source, the procedure of \aips's \texttt{jmfit} task is the logical one: calculate the S/N from the ratio of fitted peak amplitude and measured rms, and then
estimate position uncertainties as usual for correlated noise, by dividing the fitted beam sizes by the S/N, and rotate to right ascension and declination (in our case, the beam is nearly aligned, so the effects of rotation are tiny).

To derive our final positions, we used our fitting routine, taking a large $128\times128$ pixel window centered on the peak of each image to ensure a good fit.
The rms fluctuations were measured from the whole image with the $128\times128$ pixel window centered on the peak removed.
We confirmed that our fitted positions were in agreement with those from \casa\ and \aips\ and our errors were consistent with those from \aips, but different from those of \casa\footnote{Our final parallax value and uncertainty do not change if we use the \casa\ uncertainties, since the differences in the error estimates end up being absorbed by the intra-epoch errors we add later.}

From Figure~\ref{fig:VLA}, we see that both SE\_CAND3 and NE\_CAND4 are outside the FWHM of the Effelsberg beam.
To confirm that we have applied our primary beam corrections correctly and Effelsberg does not affect the positions of the candidate sources, we remove visibilities with baselines involving Effelsberg and verified that the positions remain unchanged.
As all images are calibrated to the pulsar, the positions are relative to it, and thus the inferred position of the pulsar should by definition be equal to the pointing center.
We confirmed that this was indeed the case (to well within nominal uncertainties) by fitting the Crab Pulsar's cleaned images as well.

The position uncertainties calculated this way may be slightly underestimated, since we are fitting a zero level offset instead of fixing one, and errors made in cleaning our images may not have fully propagated.
In addition, errors from EK036~B-D may be underestimated a bit more than those of EK036~A because of their sparser coverage of the $uv$ plane (EK036~A was twice as long as the other epochs and more EVN stations participated in the observation).
Finally, beyond fitting errors, there could be other residual cleaning artifacts, as well as unmodeled ionospheric and instrumental effects.

To estimate such errors for each epoch individually (``intra-epoch error''), we compare the position offsets of SE\_CAND3 inferred from the full bandwidth with offsets measured across spectral windows (similar to \citealt{Deller2009}; we omitted NE\_CAND4 as its S/N ratio in the images from the whole bands was already rather poor).
For this purpose, we made cleaned images of the sources by splitting the total bandwidth into four, $32\rm{\;MHz}$ wide parts, and fitted those to infer positions\footnote{We tried making images for every spectral window (i.e., eight $16\rm{\;MHz}$ bands) but found the S/N to be too low for reliable position measurements.}.
To account for this additional source of uncertainty, we added intra-epoch errors for SE\_CAND3 by the amount, added in quadrature to each relative position measurement in an epoch, required to produce a $\chi^{2}_{\rm red}=1$ (separately for right accession and declination; see Table~\ref{table:relative_position_fits}).
This is a somewhat more conservative approach than simply scaling the errors to obtain a $\chi^{2}_{\rm red}=1$, but ignores that with only four measurements there is a reasonable probability to find either smaller or larger $\chi^2_{\rm red}$ values by chance.
It would be worthwhile to explore this further for a larger data set.

In order to check the effect of duration, we also tried splitting the EK036~A observation in half, such that the duration and $uv$ coverage are similar to what we have in our other observations.
We find that the intra-epoch errors in both halves of the EK036~A observation increase and become comparable to those in the other observations, suggesting that increased sampling in the $uv$ plane helps minimize systematic errors.

Our final adopted positions and the associated uncertainties are listed in Table~\ref{table:relative_position_fits}.

\section{Astrometry} \label{sec:astrometry}

We use the position offsets from Table~\ref{table:relative_position_fits} to fit for the parallax ($\pi$), proper motion ($\mu_{\alpha^{\ast}}$, $\mu_{\delta}$)\footnote{We denote differences in right ascension multiplied by $\cos{\delta}$ with $\ast$.} in right ascension and declination respectively, and residual positional offset ($\Delta\alpha^\ast_{0}$, $\Delta\delta_{0}$), again in right ascension and declination respectively.
In terms of these parameters, the observed offsets are fit to,
\begin{gather}
    \begin{aligned}
        \Delta\alpha^\ast_i &= \pi f_{\alpha^\ast,i}+ \mu_{\alpha^{\ast}}(t_i-t_{0}) + \Delta\alpha^\ast_{0},\\
        \Delta\delta_i &= \pi f_{\delta,i} + \mu_{\delta}(t_i-t_{0}) + \Delta\delta_{0},
        \label{eqn:parallaxfit}
    \end{aligned}
\end{gather}
where $t_0$ is a reference time -- which we chose to be the average time over our observations (MJD 57680) to minimize covariance between the proper motion and the position offsets -- and $f_{\alpha^\ast}$ and $f_{\delta}$ are the parallax factors, given by
\begin{align}
    f_{\alpha^\ast}(t) &= X(t)\sin(\alpha_{0}) - Y(t)\cos(\alpha_{0}),
    \label{eqn:parallaxfactors2}\\
    f_{\delta}(t) &= [X(t)\cos(\alpha_{0}) + Y(t)\sin(\alpha_{0})]\sin(\delta_{0}) \nonumber\\
    &\phantom{=\ }- Z(t)\cos(\delta_{0}),
    \label{eqn:parallaxfactors1}
\end{align}
where $X(t)$, $Y(t)$, and $Z(t)$ are the components of the barycentric position of the Earth at time $t$, and $\alpha_{0}$ and $\delta_{0}$ are the approximate position of the Crab Pulsar (i.e., we neglect differences between the precise and approximate positions of the Crab Pulsar in the sine and cosine terms).
We use {\sc astropy} \citep{AstropyCollaboration2013, AstropyCollaboration2018, AstropyCollaboration2022} to calculate the barycentric positions.

As mentioned in Section~\ref{sec:obs}, SE\_CAND3 is identified also in the VLBA calibrator survey and is likely an active galactic nucleus.
We compared the differences in relative positions of SE\_CAND3 and NE\_CAND4 between the epochs.
We found these to be roughly consistent with zero and thus conclude NE\_CAND4 likely also is  extragalactic in origin.
As NE\_CAND4 is much weaker than SE\_CAND3 and its position measurements are much less reliable, we will only use SE\_CAND3 in our parallax and proper motion fits below.

\begin{figure*}
  \centering\includegraphics[width=0.985\textwidth,trim=0 0 0 0,clip]{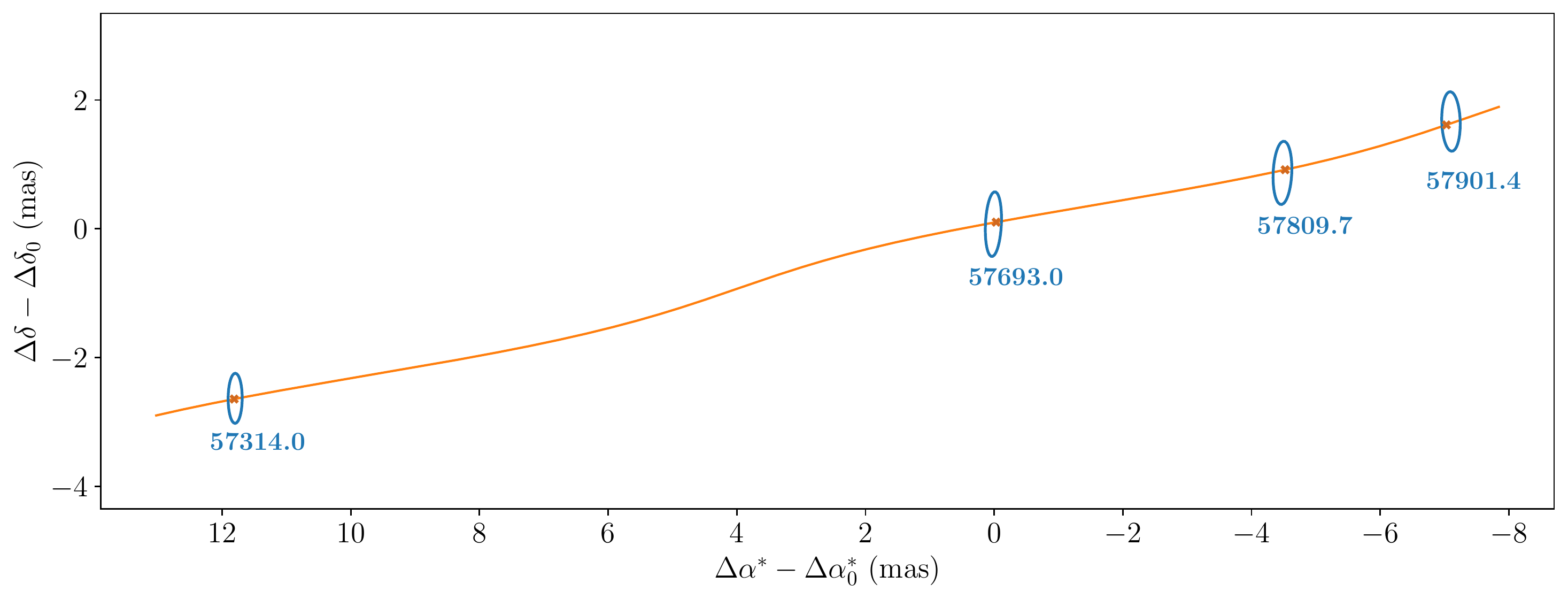}
  \caption{
    Motion of the Crab Pulsar on the sky.
    Shown are 95.4\% confidence ellipses (in blue) of our four position offsets  between the Crab Pulsar and SE\_CAND3 (see Table~\ref{table:relative_position_fits}), after subtracting the best-fit offset at our reference epoch, MJD 57680 (see Table~\ref{table:derived_parameters}).
    The modeled trajectory of the Crab Pulsar based our best-fit astrometry (see Table~\ref{table:derived_parameters}) is shown by the orange line, with the orange crosses corresponding to the modeled positions at the times of our four observations.
    \label{fig:parallax1}}
\vspace{4mm}
\end{figure*}

\begin{figure}[t]
\centering\includegraphics[width=0.47\textwidth,trim=0 0 0 0,clip]{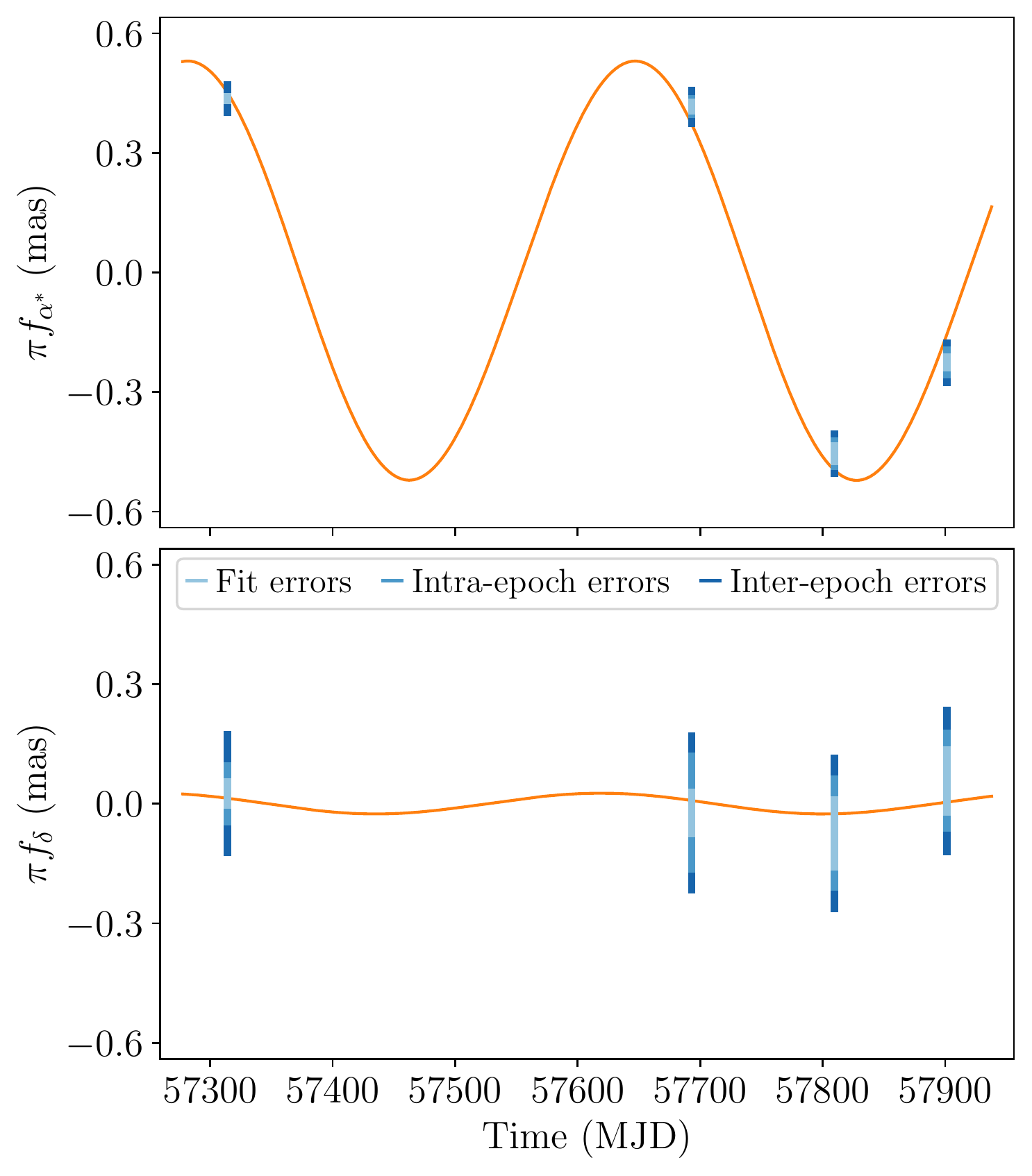}
  \caption{Position residuals in right ascension (top) and declination (bottom).
  The best-fit proper motion and relative offset between the Crab Pulsar and SE\_CAND3 have been removed, leaving just the signal due to parallax.
  Different colors in the error bars indicate the effects of the additional contributions to the uncertainties (see Table~\ref{table:relative_position_fits}).
  \label{fig:parallax2}}
\end{figure}

Our preliminary fit, including intra-epoch errors (see Section~\ref{subsec:positions} and Table~\ref{table:relative_position_fits}), yielded a parallax $\pi=0.54\pm0.03\rm{\;mas}$ and proper motion of $(\mu_{\alpha^{\ast}}, \mu_{\delta}) = (-11.31\pm0.03, 2.65\pm0.08)\rm{\;mas\;yr^{-1}}$.
We find $\chi^{2}_{\rm red}=2.3$, larger than the expected unity.
This could simply reflect that we have very few degrees of freedom: in particular, the parallax fit is dominated by the four right ascension offsets, to which three parameters are fitted, leaving only a single degree of freedom.
Indeed, \citet{Reid2017} showed that with four epochs and one effective degree of freedom, the uncertainty on the uncertainty in the parallax can be significant.
Still, we will assume conservatively that, instead, there are unmodeled systematic errors between epochs (``inter-epoch errors'').
We estimate these at $0.04\rm{\;mas}$ and $0.14\rm{\;mas}$, for right accession and declination respectively, the value that, added in quadrature to the measurement errors of both right ascension and declination in all epochs, gives a $\chi^{2}_{\rm red}=1$ (see Table~\ref{table:relative_position_fits}).
The inter-epoch errors in right ascension and declination were taken to be roughly proportional to the beam size, as might be expected if the systematic effects are due to phasing errors\footnote{Our results suggest the error in declination may be overestimated. If we take errors that are the same in each coordinate, we find we require these to be $0.04{\rm\;mas}$.
With these, we find identical results except for a somewhat reduced final error in the proper motion in declination.}.
With these, we derive the final fit results presented in Table~\ref{table:derived_parameters} and shown in Figures~\ref{fig:parallax1}~and~\ref{fig:parallax2}.

\begin{deluxetable}{lrr}[t]
\tabletypesize{\small}
\setlength{\tabcolsep}{2.75pt}
\tablecaption{Astrometric Parameters\label{table:derived_parameters}}
\tablenum{5}
\tablehead{
  \colhead{Parameter}&
  \colhead{EVN}&
  \colhead{\emph{Gaia} DR3}
}
%% All data must appear between the \startdata and \enddata commands
\startdata
$\pi\rm{\;(mas)}$\dotfill                        & $0.53\pm0.06$                               & $0.51\pm0.08$          \\
$\mu_{\alpha^{\ast}}\rm{\;(mas\;yr^{-1})}\ldots$ & $-11.34\pm0.06$                             & $-11.51\pm0.10$        \\
$\mu_{\delta}\rm{\;(mas\;yr^{-1})}$\dotfill      & $2.65\pm0.14$                               & $2.30\pm0.06$          \\
$\alpha_{\rm J2000}$\dotfill                     & $5^{\rm{h}}34^{\rm{m}}31.93357^{\rm{s}}$ & $5^{\rm{h}}34^{\rm{m}}31.933561(5)^{\rm{s}}$ \\
$\delta_{\rm J2000}$\dotfill                     & $22\arcdeg00\arcmin52.1927\arcsec$       & $22\arcdeg00\arcmin52.19236(6)\arcsec$ \\
$d\rm{\;(kpc)}$\dotfill                          & $1.90^{+0.22}_{-0.18}$                      & $1.96^{+0.36}_{-0.26}$ \\
$v_{\perp}\rm{\;(km\;s^{-1})}$\dotfill           & $104^{\phantom{.0}+13}_{\phantom{.0}-11}$   & $109^{\phantom{.0}+21}_{\phantom{.0}-15}$ \\
\enddata
\tablecomments{
Shown are both our results and those from \emph{Gaia}.
Distances are calculated directly from the parallax measurements and the transverse velocity $v_{\perp}$ is inferred from the proper motion and inferred distance.
Coordinates listed here are in the J2000 ICRS frame at MJD 57680 (our reference epoch), with the uncertainties in our EVN results dominated by the uncertainty in the position of our reference source ($\sim\!1\rm{\;mas}$, see text), and those for \emph{Gaia} given by the values in parentheses.}
\end{deluxetable}

\vspace{-4mm}
We also split the EK036~A observation in half in time and use the source position fits obtained from each half as independent measurements in a new fit for the parallax and proper motion.
We find no significant changes in our fit parameters; however, the error on the parallax reduces a little and there is less of a need for an inter-epoch contribution.
Since this may just be a statistical fluke, we continue with our regular solution below.

One possible cause of systematic errors between epochs might be residual ionospheric errors between the pulsar and SE\_CAND3.
To give a sense of the size of the error from differences between the mean path length through the ionosphere, we find from CDDIS TEC maps that the average residual vertical TEC between antennas for SE\_CAND3 relative to the pulsar is $\sim\!0.02\rm{\;TECU}$.
Though the resolution and accuracy of the TEC maps are poor, if we take the residual vertical TEC at face value, this translates to an extra path length of $\sim\!0.3\rm{\;cm}$, which, if systematic over all telescopes, might induce position offsets of up to $\sim\!0.06\rm{\;mas}$, comparable to the inter-epoch errors we infer.
Indeed, in Appendix~\ref{sec:appendix}, we show that the inclusion of residual ionosphere correction from TEC map information results in shifts in position of this order of magnitude.
We also find this leads to somewhat smaller inferred inter-epoch error and thus smaller uncertainties in the astrometric parameters, but do not feel confident enough in these results to use them (see Appendix~\ref{sec:appendix}).

Another source of systematic error may come from refraction in the interstellar medium.
This will affect both the calibrators and the pulsar, but differently.
For an estimate, we use that \citet{Rudnitskii2016} measured an scattering disk with full width at half maximum varying between $0.5\!-\!1.3\rm{\;mas}$ at $18\rm{\;cm}$.
The variability suggests that at times the screen is asymmetric, which would lead to position offsets if not accounted for.
If this induces relative position shifts of order 10\% of the width, which seems not unreasonable, it would induce offsets of $\sim\!0.05\rm{\;mas}$, the right order of magnitude to account for the possible systematic errors between epochs.

\begin{figure}
\centering
\includegraphics[width=0.47\textwidth,trim=0 0 0 0,clip]{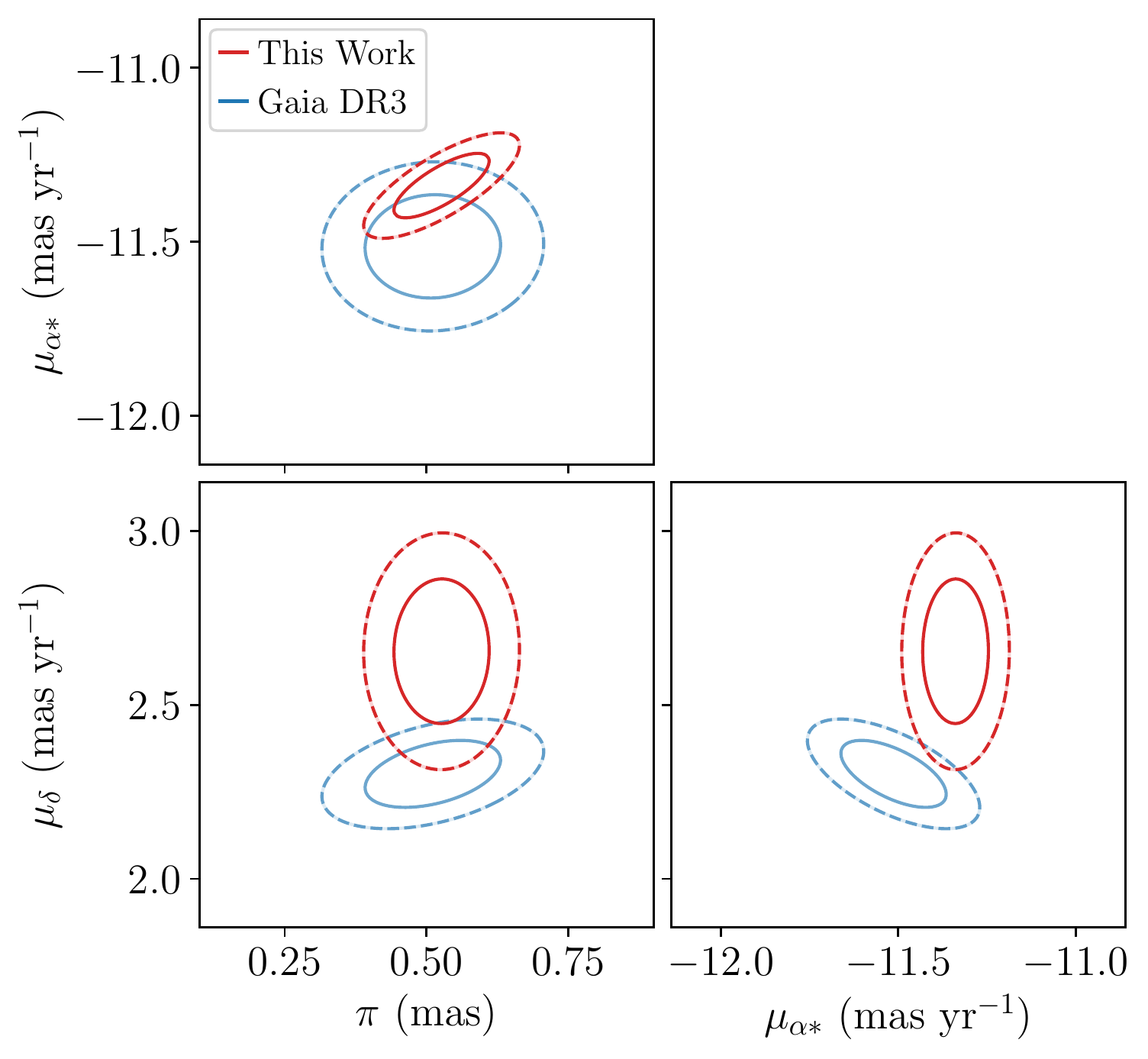}
\caption{Error ellipses of the parallax and proper motion from our work (in red) and \emph{Gaia} DR3 (in blue).
The 68.3\% and 95.4\% confidence ellipses are shown by the solid and dashed ellipses, respectively. \label{fig:confidence_ellipse}}
\end{figure}

Finally, in our source images we see no apparent jets or other structures that could induce positional errors.
However, we note that \citet{Koryukova2022} found that for SE\_CAND3, the measured angular core size appeared to vary between $\sim\!0.07$ and $1.56\rm{\;mas}$ at $4-8\rm{\;GHz}$ for two observations separated by $2.6\rm{\;yr}$.
If real, this variability might also change the centroid by amounts comparable to the systematic errors we infer between epochs.

\section{Results} \label{sec:results}

We measure a parallax of $\pi=0.53\pm0.06\rm{\;mas}$ for the Crab Pulsar and infer a distance of $d=1.90^{+0.22}_{-0.18}$ by taking the reciprocal of the measured parallax (we do not attempt to correct for \cite{Lutz1973} bias, as it is not clear what the prior likelihood of finding a supernova remnant at a given height above the galactic plane would be).
From our best-fit proper motion and inferred distance, we also derive a transverse velocity of $v_{\perp}=104^{+13}_{-11}\rm{\;km\;s^{-1}}$.
Using the coordinates of SE\_CAND3 (WFCS~J0535+2156; \citealt{Petrov2021}),
\begin{align}
    \begin{aligned}
        \alpha_{\rm J2000} &= 5^{\rm{h}}35^{\rm{m}}06.34125^{\rm{s}},\\
        \delta_{\rm J2000} &= 21\arcdeg56\arcmin49.1045\arcsec,
    \label{eqn:secand3_position}
    \end{aligned}
\end{align}
in the J2000 International Celestial Reference System (ICRS) frame, we determine the absolute position of the Crab Pulsar, in the same reference frame, at MJD 57680, as,
\begin{align}
    \begin{aligned}
        \alpha_{\rm J2000} &= 5^{\rm{h}}34^{\rm{m}}31.93357^{\rm{s}},\\
        \delta_{\rm J2000} &= 22\arcdeg00\arcmin52.1927\arcsec.
    \label{eqn:crab_position}
    \end{aligned}
\end{align}
The uncertainty in our position for the Crab Pulsar is dominated by the uncertainty in the position of SE\_CAND3.
The formal errors are $\pm0.6\rm{\;mas}$ in each coordinate, but those are for the positions measured at $4\!-\!8\rm{\;GHz}$ and we have not accounted for possible frequency dependent core-shifts, which typically are of order $1\rm{\;mas}$ \citep{Sokolovsky2011}.
Hence, we estimate the uncertainties in the position at $\sim\!1\rm{\;mas}$ in each coordinate.
Our measured and derived values are presented in Table~\ref{table:derived_parameters}.

Comparing our results with those of \emph{Gaia} DR3, listed also in Table~\ref{table:derived_parameters}, we find good agreement for the parallax but some tension for the proper motion.
To investigate this further, we show confidence ellipses of our parallax and proper motion along with those from \emph{Gaia} DR3 in Figure~\ref{fig:confidence_ellipse}.
One sees that the main discrepancy is for the proper motion in declination.
Our measurements are less sensitive in declination, since most EVN telescopes are spread East-West, with most of the North-South constraint coming from Hartebeesthoek.
Thus, we may still underestimate the uncertainty of the proper motion in declination.
Fortunately, this should not affect the parallax: since the Crab Pulsar is near the ecliptic, the parallax barely correlates with the proper motion in declination.
It has some correlation with proper motion in right ascension, and, taking our error ellipse and that of \emph{Gaia} DR3 at face value, a slightly lower parallax might be inferred.

We note that systematic effects may affect not just our measurement (see above), but also the \emph{Gaia} DR3 astrometry of the Crab Pulsar.
Indeed, the values for the proper motion presented in \emph{Gaia} DR2 and DR3 differ significantly (see Section~\ref{sec:intro}).
For the parallax, there is a possible overall zero-point correction, but this is a small effect: applying the correction of $-0.03\rm{\;mas}$ from \cite{Lindegren2021a} to the raw \emph{Gaia} DR3 parallax from Table~\ref{table:derived_parameters}, yields $\pi=0.54\pm0.08\rm{\;mas}$ (and an inferred distance of $d=1.86_{-0.24}^{+0.32}$), which still agrees well with our measured and inferred results.
Another possible systematic effect is due to source color.
In \emph{Gaia} DR3, a 6-parameter fit including the pseudo-color was used for the astrometry, and the solution shows fairly strong covariance between the pseudo-color and the proper motion.
According to \cite{Lindegren2021b}, for cases where strong correlation is seen, independent colour information may significantly improve precision and accuracy.
Here, one would have to be somewhat careful, since the Crab Pulsar's spectrum is not like that of regular stars, for which the color corrections are calibrated.
Finally, it also seems possible that the variable optical emission surrounding the Crab Pulsar, such as the wisp-like structures moving outwards from the pulsar, and halos and knots close to it \citep{Hester2002, Tziamtzis2009}, might induce positional offsets that could affect the astrometry.
We conclude that in both optical and radio it will be useful to analyze further observations and try to carefully account for potential biases and systematic effects.

\section{Future Work} \label{sec:futurework}

Our pilot study shows that it is possible to measure the parallax of the Crab Pulsar with VLBI.
It should be relatively straightforward to improve the measurement down to the $\lesssim\!5\%$ level with further VLBI observations.
The EVN's extended East-West baseline is particularly useful for constraining the parallax of the Crab Pulsar as the synthesized beam is narrower in right ascension and the pulsar is very close to the ecliptic.
More observations should be scheduled around October and March when the parallax signature would peak in right ascension.

Future observations should try to include more small dishes to give maximum sensitivity for the in-beam extragalactic reference sources.
Furthermore, the pointing center can be shifted towards SE\_CAND3 and NE\_CAND4 (e.g. centroid of all sources) to boost the signal of those sources.
With the higher sensitivity, NE\_CAND4 should become more useful in helping to constrain and verify the astrometry.
The addition of NE\_CAND4 may also allow one to use the MultiView technique \citep{Rioja2017}, or variants thereof (e.g. \citealt{Reid2017, Hyland2022}), which has shown success in improving residual spatial ionospheric corrections.

We have shown that our technique of using giant pulses to determine fringe and bandpass solutions works exceedingly well for self-calibration, removing the need to observe phase calibrators.
Our estimates of systematic effects between epochs suggests it is better to have a larger number of observations rather than to have longer ones.
However, since the intra-epoch error for EK036~A is quite a bit smaller than in EK036~B-D, one would not want to reduce the time too much.

With more observations and better time coverage, the error analysis could be improved, e.g., using a bootstrap fit like was done by \cite{Deller2019}.
Overall, we suggest at least 8 to 9 observations, each lasting at least $2\rm{\;hr}$ in order to ensure sufficient $uv$ coverage.
If the detection rate of strong giant pulses remain high enough for self-calibration, it may be better to observe at slightly higher frequencies, say $\sim\!2\rm{\;GHz}$, to reduce the effects of ionospheric variations and interstellar scattering.
Calibration of both scattering and residual ionospheric effects would be helped by simultaneous dual-frequency or wide-band ($\gtrapprox\!350\rm{\;MHz}$) observations \citep{Brisken2000,Petrov2023}.
These wider-band observations may allow for an alternative measurement of the small differences in contributions from the ionosphere between the Crab Pulsar and the in-beam calibrators and improve on the application of TEC maps described in Appendix~\ref{sec:appendix}.
Of course, ionospheric errors can also be reduced by trying to schedule observations when the solar cycle is at its minimum.

Our technique of self-calibration using giant pulses should also help future studies of the Crab Pulsar's environment, such as the flaring regions within the Crab Nebula studied by \cite{Lobanov2011}.
Furthermore, the technique may also be useful for measuring distances to other giant pulse emitters such PSR~J1824$-$2452A \citep{Bilous2015} and PSR~J1823$-$3021A \citep{Abbate2020}, as well as to bright rotating radio transients\footnote{\url{http://astro.phys.wvu.edu/rratalog/}} such as PSR~J1819$-$1458 and PSR~J1840$-$1419, which have bursts every $\sim\!3.4\rm{\;min}$ and $\sim\!1.3\rm{\;min}$, respectively \citep{McLaughlin2006}.
For PSR~J1824–2452A and PSR~J1823-3021A, which are both in globular clusters, using their pulses for phase calibration would also aid searches of further globular cluster pulsars and other radio emitters.
Similarly, applying our technique to the Crab Pulsar twin PSR~J0540-6919, which also exhibits giant pulses \citep{Geyer2021}, may help searches of new radio sources in the Large Magellanic Cloud.

Our cleaned images in FITS format are made available as a dataset at\dataset[10.5281/zenodo.7910778]{\doi{10.5281/zenodo.7910778}}.
The raw baseband data along with our custom scripts are available upon request\footnote{As the baseband data were correlated by us and not by the JIVE team, the visibility products are not available on the EVN Data Archive.}.

\section*{Acknowledgements}
We thank Cees Bassa for his contribution to the EVN proposal and the anonymous referee for useful comments.
R.L. thanks Aard Keimpena, Bob Campbell, Benito Marcote, and Marjolein Verkouter for useful advice on using \sfxc\ and JIVE post-processing tools. R.L. thanks the National Radio Astronomy Observatory (NRAO) helpdesk for advice on using \casa, including details of the structure of the various tables, and for providing access to the NRAO computing facilities.
Computations were performed on the New Mexico Array Science Center (NMASC) cluster and the Niagara supercomputer at the SciNet HPC Consortium \citep{Loken2010, Ponce2019}.
SciNet is funded by: the Canada Foundation for Innovation; the Government of Ontario; Ontario Research Fund - Research Excellence; and the University of Toronto.
M.Hv.K. is supported by the Natural Sciences and Engineering Research Council of Canada (NSERC) via discovery and accelerator grants, and by a Killam Fellowship.
F.K. acknowledges support from the Onsala Space Observatory for the provisioning of its facilities/observational support.
The Onsala Space Observatory national research infrastructure is funded through Swedish Research Council grant No 2017-00648.
U.-L.P. receives support from Ontario Research Fund-Research Excellence Program (ORF-RE), NSERC [funding reference Nos. RGPIN-2019-067, CRD 523638-18, 555585-20], Canadian Institute for Advanced Research (CIFAR), the National Science Foundation of China (grant No. 11929301), Alexander von Humboldt Foundation, and the National Science and Technology Council (NSTC) of Taiwan (111-2123-M-001, -008, and 111-2811-M-001, -040).

\facilities{The European VLBI Network is a joint facility of independent European, African, Asian, and North American radio astronomy institutes. Scientific results from data presented in this publication are derived from the following EVN project codes: EK036~A-D.
The NRAO is a facility of the National Science Foundation operated under cooperative agreement by Associated Universities, Inc.}

\software{
  astropy \citep{AstropyCollaboration2013, AstropyCollaboration2018, AstropyCollaboration2022},
  Baseband \citep{VanKerkwijk2020},
  CALC10 \citep{Ryan1980},
  CASA \citep{CASA2022},
  numpy \citep{Harris2020},
  matplotlib \citep{Hunter2007},
  pulsarbat \citep{Mahajan2022},
  scipy \citep{Gommers2022},
  SFXC \citep{Keimpema2015},
  tempo2 \citep{Hobbs2012}.}

\appendix
\vspace{-4mm}
\section{Differential Ionosphere Corrections using TEC Maps}\label{sec:appendix}

A source of error in our position measurements of the Crab Pulsar relative to our in-beam calibrators arises from slight differences in the total electron column (TEC) in the ionosphere between the different sources.
Estimates of these differences can be made from TEC maps, as is becoming common in VLBI astrometry.
While this use of TEC maps, including the underlying assumptions about the ionosphere, have not been fully validated (see \citealt{Petrov2023}), we follow it to get a sense of the improvement that may be attainable.
Since our giant-pulse based fringe solutions already include the contribution of the ionosphere towards the Crab Pulsar (along with delays introduced by antenna location, electronics, geometric models, etc.), we only need to apply a differential correction for the extragalactic sources.

To determine the residual ionospheric corrections, we first download CDDIS TEC maps using \casa's \texttt{tec\_maps} function.
We then use \casa's \texttt{gencal} task to estimate the line-of-sight TEC from each antenna to each of our sources across each observation (\texttt{gencal} models the ionosphere as a thin shell at a constant height of $450\rm{\;km}$).
Using custom scripts, we then calculate the differential TEC between the Crab Pulsar and extragalactic sources for each antenna and write the residuals into our own \casa\ compatible calibration tables.
These new calibration tables are applied to the visibilities data of SE\_CAND3 and NE\_CAND4 using \casa's \texttt{applycal} task (after applying the calibrations described in Section~\ref{subsec:calibration}).
We then create images and extract position offsets as in Sections~\ref{subsec:imaging} and \ref{subsec:positions}.
We list the resulting offsets in Table~\ref{table:relative_position_fits_tec}, and fit these  to our astrometric model (including intra- and inter-epoch errors estimated like in Sections~\ref{subsec:positions} and \ref{sec:astrometry}).

We find a parallax of $\pi=0.49\pm0.04\rm{\;mas}$ and proper motion of $(\mu_{\alpha},\mu_{\delta})=(-11.41\pm0.05,2.54\pm0.11)\rm{\;mas\;yr^{-1}}$, i.e., values consistent with our results in Section~\ref{sec:results} and with \emph{Gaia} DR3.
We note that the uncertainties are slightly reduced, a consequence of the fit to the offsets being somewhat better, thus reducing the estimated inter-epoch error contribution to the uncertainties.
While encouraging, we caution that with the small number of data points, a reduction by chance is not unlikely, in particular in the presence of possible other sources of systematic error such as refraction in the interstellar medium and source variability (see Section~\ref{sec:astrometry}).

As a further check on the reliability, we also tried applying TEC corrections when transferring calibrator solutions to the pulsar as above, but this time applying a differential correction for the pulsar.
As the angular separations of the calibrator sources and pulsar are quite large and the calibrator/pulsar cycle is quite long, we also tried removing the ionospheric contributions towards the calibrators using the TEC maps before solving for the calibrator fringes (in the hopes that these new calibrator fringe solutions with slower time variations can be better extrapolated to the pulsar).
We then applied TEC corrections towards the Crab Pulsar and the new calibrator fringe solution to the pulsar. Both methods resulted in similar quality images.
If the corrections were good, we expect that with these solutions, the dirty images would improve, i.e., that we would see the Crab Pulsar becoming more point-like.
However, we found that with the TEC corrections, the dirty images were of poorer quality (more smeared) than those shown in the top panels of Figure~\ref{fig:dirty_crab_images}.
Given this contradictory result, we concluded that without better understanding it was best not to use the above TEC-map assisted astrometry, even though it gave notionally better results.
Since our ``ionosphere corrected'' SE\_CAND3 and NE\_CAND4 images may still be useful for future astrometry of the Crab Pulsar, we provide these (along with those from Figure~\ref{fig:clean_images}) at \dataset[10.5281/zenodo.7910778]{\doi{10.5281/zenodo.7910778}}.

\begin{deluxetable}{ccccc}[t]
\tabletypesize{\small}
\setlength{\tabcolsep}{6pt}
\tablecaption{Relative Positions between the Reference Sources and the Crab Pulsar, with Ionosphere Corrections applied.\label{table:relative_position_fits_tec}}
\tablenum{A.1}
\tablehead{\colhead{Observation}&
\multicolumn{2}{c}{\dotfill SE\_CAND3 \dotfill}&
\multicolumn{2}{c}{\dotfill NE\_CAND4 \dotfill}\\[-.7em]
\colhead{code}&
\colhead{$\Delta\alpha^{\ast}$ ($\rm{mas}$)}&
\colhead{$\Delta\delta$ ($\rm{mas}$)}&
\colhead{$\Delta\alpha^{\ast}$ ($\rm{mas}$)}&
\colhead{$\Delta\delta$ ($\rm{mas}$)}}
%% All data must appear between the \startdata and \enddata commands
\startdata
EK036~A & $-478472.634\pm0.015\pm0.00\pm0.03$            & $243085.61\pm0.04\pm0.08\pm0.09$ & $-585690.03\pm0.14$           & $-195268.3\pm0.3$\\
EK036~B & $-478484.582\pm0.02\phantom{0}\pm0.02\pm0.03$  & $243088.27\pm0.06\pm0.14\pm0.09$ & $-585702.12\pm0.2\phantom{0}$ & $-195265.3\pm0.5$\\
EK036~C & $-478489.033\pm0.03\phantom{0}\pm0.02\pm0.03$  & $243088.92\pm0.09\pm0.10\pm0.09$ & $-585706.43\pm0.18$           & $-195264.2\pm0.4$\\
EK036~D & $-478491.501\pm0.02\phantom{0}\pm0.03\pm0.03$  & $243089.78\pm0.09\pm0.09\pm0.09$ & $-585708.26\pm0.19$           & $-195261.8\pm0.5$\\
\enddata
\tablecomments{Values and uncertainties are as for Table~\ref{table:relative_position_fits}, except that here they were derived from data for which we tried to correct for differential ionospheric effects using TEC maps.}
\end{deluxetable}
\vspace{-4mm}

\bibliographystyle{aasjournal}
\bibliography{main}

\end{document}